\documentclass[a4paper,fleqn]{cas-sc}

\usepackage[authoryear,longnamesfirst]{natbib}

\usepackage{subfigure}
\usepackage{lineno}
\usepackage{comment}
\usepackage{color}

\usepackage{multirow}
\usepackage{booktabs}

\usepackage{adjustbox}

\usepackage[linesnumbered,ruled,vlined]{algorithm2e}

\usepackage{pifont}
\def\tsc#1{\csdef{#1}{\textsc{\lowercase{#1}}\xspace}}
\tsc{WGM}
\tsc{QE}
\tsc{EP}
\tsc{PMS}
\tsc{BEC}
\tsc{DE}
\begin{document}
\let\WriteBookmarks\relax
\def\floatpagepagefraction{1}
\def\textpagefraction{.001}

\shorttitle{Multi-Factor Authentication Threat Modelling Survey}

\shortauthors{Bezerra, Souza, Westphall and Westphall}

\title [mode = title]{Characteristics and Main Threats about Multi-Factor Authentication: A Survey}                      
%\tnotemark[1,2]

%\tnotetext[1]{This document is the results of the research
%   project funded by the National Science Foundation.}

%\tnotetext[2]{The second title footnote which is a longer text matter
%   to fill through the whole text width and overflow into
%   another line in the footnotes area of the first page.}

\author[1,3]{WR Bezerra}[%type=Editor,
                        %auid=000,bioid=1,
                        role=Phd Candidate, orcid=0000-0002-6098-7172]
\cormark[2]
\fnmark[1]
\ead{wesleybez@gmail.com}

\author[1,3]{CA de Souza}[%type=Editor,
                        role=Phd Candidate, orcid=0000-0002-9453-3240]
\cormark[2]
\fnmark[1]
\ead{cristianoantonio.souza10@gmail.com}

\author[1,3]{CM Westphall}[%type=Editor,
                        role=Phd, orcid= 0000-0002-7213-1603]
\cormark[2]
\fnmark[1]
\ead{carla.merkle.westphall@ufsc.br}

\author[1]{Carlos B. Westphall}[role=Advisor,
        orcid=0000-0002-5391-7942]
\cormark[1]
\ead{carlosbwestphall@gmail.com}

\address[1]{UFSC - Federal University of Santa Catarina,Campus Universitário - Trindade,Florianópolis/SC - Brazil, 88040-380}

\cortext[cor1]{Corresponding author}
\cortext[cor2]{Principal corresponding author}

%\fntext[fn1]{This is the first author footnote. but is common to third
%  author as well.}
%\fntext[fn2]{Another author footnote, this is a very long footnote and
%  it should be a really long footnote. But this footnote is not yet
%  sufficiently long enough to make two lines of footnote text.}

%\nonumnote{This note has no numbers. %In this work we demonstrate $a_b$
%  the formation Y\_1 of a new type of polariton on the interface
%  between a cuprous oxide slab and a polystyrene micro-sphere placed
%  on the slab.
%  }

\begin{abstract}
    This work reports that the Systematic Literature Review process is responsible for providing theoretical support to research in the Threat Model and Multi-Factor Authentication. However, different from the related works, this study aims to evaluate the main characteristics of authentication solutions and their threat model. Also, it intends to list characteristics, threats, and related content to a state-of-art. As a result, we brought a portfolio analysis through charts, figures, and tables presented in the discussion section. 
\end{abstract}

% \includegraphics{figs/grabs.pdf}

\begin{highlights}
\item a document portfolio for threat model and multi-factor authentication areas that represents the areas' state-of-art;

\item a list of the main characteristics of multi-factor authentication researches from the portfolio;

\item a list of the main threats to multi-factor authentication obtained from the state-of-art.
\end{highlights}

% Keywords

\maketitle

\section{Introduction}
\label{S:1}

This work reports that the Systematic Literature Review process is responsible for providing theoretical support to research in Threat Model (TM) and Multi-Factor Authentication (MFA). Also, it aims to build a bibliographic portfolio capable of guiding the discussions and being theoretical support necessary for putting forward the previously cited research areas. 

The present work focuses on multi-factor authentication, a sub-area of authentication that is wide-ranging. Some related works in this sub-area ranges from authentication schemes \cite{chaudhry2019improved}, biometrics storage \cite{ali2018edge}, the diversity between authentication factors \cite{loffi2021mutual,anakath2019privacy,ometov2018multi}. However, different from such works, the documents selected evaluate the main characteristics of authentication solutions and their threat model.

Therefore, this work focuses on \textit{threat models} for multi-factor authentication. It intends to list characteristics, threats, and related content to state-of-art research in MFA. As a result, white papers, patents, or less academic documents were removed from the portfolio selection. In general, this review aims to answer the following research questions:
\begin{itemize}
    \item[Q1] what are the \textbf{main articles} in the selected research area?
    \item[Q2] what are the \textbf{main characteristics} intended by the analyzed authentication solutions?
    \item[Q3] what are the \textbf{main threats} listed in the \textit{threat models} that appear in the selected portfolio?
\end{itemize}

The work continues with the Systematic Literature Review in section \ref{S:2}; the sequence brought a discussion about the results in \ref{S:3}. In the \ref{S:4}, the conclusions and future work for this section are brought.

\begin{table}[t]
    \centering
    \caption{List of Abbreviations}
    \begin{tabular}{ll}
        \toprule
        Abbreviation & Meaning  \\
        \toprule
        CR  & Challenge-Response\\
        DDoS& Distributed Denial of Service\\
        DFD & Data Flow Diagram\\
        MFA & Multi-Factor Authentication\\
        OTP & One-Time Password\\
        SLR & Systematic Literature Review\\
        SSO & Single Sign-On\\
        TM  & Threat Model\\
        U2F & Universal 2nd Factor\\
        \bottomrule
    \end{tabular}
    \label{tab:abbreviations}
\end{table}

\section{Systematic Literature Review (SLR)}
\label{S:2}

The systematic literature review is a process/methodology that promotes the reduction of bias in scientific research \cite{kitchenham2009systematic}. However, it is not limited to this type of research \cite{moher2009preferred,budgen2006performing}, but expands its results to the construction of didactic material, classes, and books, being a solid base for building a knowledge base.

In this work, the adapted ProKnow-C \cite{ensslin2015outsourcing} methodology was used as the Systematic Literature Review methodology. Such methodology consists of four \textit{macrosteps}: (i) portfolio selection, (ii) systematic review, (iii) bibliometrics, and (iv) research questions. Also, the research questions were already established \textit{a priori}, and the answer to the research questions replaced the last step.

Therefore, in this section, the macrosteps of (i) portfolio selection and (ii) systematic review will be explored. The portfolio was chosen through these macro-steps in a documented and replicable manner. Additionally, the lenses (points of view) were chosen, analyzed, and contributed to the successful conclusion of this study. Thus, this review phase is the most important in this work.

%string de consulta usada
\begin{equation}
 \begin{split}
&\textquotesingle threat model\textquotesingle\;  AND\; \textquotesingle multi-factor\; authentication\textquotesingle \\
&AND\;  PUBYEAR\;  >\;  2018  \\
&AND\;  ( LIMIT-TO ( DOCTYPE ,  \textquotesingle ar\textquotesingle )  \\
&OR\;  LIMIT-TO ( DOCTYPE ,  \textquotesingle cp\textquotesingle )  \\
&OR  LIMIT-TO ( DOCTYPE ,  \textquotesingle re\textquotesingle ) )  
    \end{split}    
    \label{eq:query_slr}
\end{equation}

The files used during the systematic review process are available at the following link\footnote{https://github.com/wesleybez/mfar\_tm}.

\subsection{Portfolio Selection}
\label{S:2.2}
%banco de dados de artigos inicialmente com 34 publicações

This macrostep is a systematization for selecting articles that comprise this study's final portfolio of base articles. This portfolio well represents the research object, the chosen cut, and the purpose of this cut in the research. Therefore, concise but representative of the state-of-the-art in the researched area.

In this SLR, the research object is state-of-the-art Threat Models for Multi-Factor Authentication. Thus, the object of study can be translated into the following query-string (\ref{eq:query_slr}), which is limited to the last four years (since 2018) and applies to articles (ar), reviews (re), and conference papers (cp).

\begin{table}[h!]
    \centering
    \caption{Inclusion and Exclusion Criteria for selection of works - i$X$ or e$X$, are used for indexing the criteria, where: the prefix i designates inclusion criteria, the prefix e designates exclusion criteria, and the $X$ must be replaced by the count of the number of criteria in each case.}
    \begin{tabular}{p{0.15\textwidth}p{0.6\textwidth}}
    \toprule
        i1 & documents published from 2018 \\
        i2 & documents in English \\
    \midrule
        e1 & \textit{impact factor} less than 1.0 \\
        e2 & repeated work report \\
        e3 & document not available in full \\
    \bottomrule
    \end{tabular}
    
    \label{tab:criteria_inc_exc}
\end{table}

The inclusion and exclusion criteria from Table \ref{tab:criteria_inc_exc} were used as a guide for selecting works. As for inclusion, criteria (i1) can be mentioned regarding the maximum age of the articles being four years old, and (i2) regarding the articles reporting the research in English. As for the exclusion criteria, criterion (e1) indicates removing articles in sources that have the \textit{impact factor}\footnote{\textit{Impact factor} obtained in the Scimago Journal \& Country Rank} lower than 1.0, the (e2) indicates that repeated articles or that are a repeated report by the same researcher but in a different source are excluded, and (e3) that regulates that works that do not have full access to their content are removed from the analysis flow. Therefore, these criteria (Table \ref{tab:criteria_inc_exc}) helped to make a more deterministic selection and to avoid bias.

The portal chosen for the present work was SCOPUS. This portal is accessible through the CAPES Periodic agreement, and it is possible to access most of the articles published therein from the public higher education system. Furthermore, this portal allows us to \textit{download} a database of research articles containing many fields relevant to the research. Thereupon, this database was sufficient due to it incorporates a vast number of sources from different publishers in computer science.

\begin{table}[h!]
    \centering
    \caption{Description of the analyzed articles database. As for data types: (i) integer, (r) real, (t) text.}
    \begin{tabular}{lll}
        \toprule
        field & type & description \\
        \midrule
        index & i & \makecell[l]{sequential number used to identify the article in the\\ workflow} \\
        title & t &  field with document title \\
        source & t & field with the name of the source where the document was published\\
        citations & i & field that stores the number of citations of the work\\
        impact factor & r & \makecell[l]{stores the impact factor value of the source where the\\ article was published}\\
        hindex & i & \makecell[l]{stores the h-index value of the source where the article \\was published}\\
        year & i & stores the year of publication of the article\\
        \bottomrule
    \end{tabular}
    
    \label{tab:article_database_description}
\end{table}

After the search, 32 articles were obtained for the initial database. Such articles were registered using a spreadsheet with fields already detailed in Table \ref{tab:article_database_description}. The spreadsheet management tool LibreOffice Calc\footnote{https://pt-br.libreoffice.org/descubra/calc/} was used to tab the data and generate the datasets. For document management, the tool Mendeley\footnote{https://www.mendeley.com/search/} was used, which allows the organization of documents in folders and subfolders that replicate the portfolio selection steps. This tool considerably facilitated the work of organizing the documents in stages.

The previously mentioned inclusion and exclusion criteria can be applied to the database with the initial database and the data from the tabulated articles. Thus, at the end of the \textbf{initial phase}, 27 articles remained, and of the seven excluded articles (e1), two did not have the impact factor \cite{roy2017cryptanalysis,de2018secret}, four \cite{simpson2017enterprise,simpson2018insider,hawrylak2019practical,kara2020sauglik} had an impact factor less than 1.0, and one did not found\footnote{Search performed on Google Scholar}.

Following the workflow, seven more articles \cite{johansen2014probabilistic,ma2016openid,gonzalez2019formalizing,chen2020risk,gupta2020machine,lu2021efficient,gonzalez2021formalizing} were removed in the \textbf{title reading phase}, leaving only 20 articles for the next phase. Of these, another seven were excluded during the \textbf{abstracts reading phase}, leaving only 13 articles for the complete reading phase. The document used for the decision to exclude articles based on reading the \textit{abstracts} can be seen at the link\footnote{https://github.com/wesleybez/mfar\_tm} of the project on GitHub.

After the \textbf{complete reading phase}, four articles were excluded, of which: one was a duplicate publication in different \textit{sources} but by the same authors; the complete document was chosen \cite{jacomme2021extensive}; the following article removed did not have free access to read the content of the work in its entirety \cite{sinigaglia2019mufasa}; the next one presented a very broad approach to the need for the present study \cite{mahbub2020progressive}; and the last \cite{bojjagani2021systematic} diverged from the focus of the study, having as its specific object the study of \textit{mobile payment} and not \textit{authentication}. In this way, nine articles were left composing the bibliographic portfolio of this work.

\begin{table}[h!]
    \centering
    \caption{Systematic bibliographic portfolio. This article set represents the selection made after the various stages of analysis performed on the set resulting from the initial search.}
     \begin{adjustbox}{max width=.85\textwidth}
    \begin{tabular}{lll}
    \toprule
        \# & Reference & Title\\
    \midrule
        a01 & \cite{chen2012mobile} & \makecell[l]{Mobile device integration of a fingerprint biometric remote\\ authentication scheme}\\ 
        \midrule
        a02 & \cite{li2014unified}  & \makecell[l]{Unified threat model for analyzing and evaluating software\\ threats}\\
        \midrule
        a05 & \cite{dhillon2017secure} & \makecell[l]{Secure multi-factor remote user authentication scheme for\\ Internet of Things environments}\\
        \midrule
        a13 & \cite{ferrag2019authentication} & \makecell[l]{Authentication and Authorization for Mobile\\ IoT Devices Using Biofeatures: Recent Advances and Future Trends}\\
        \midrule
        a17 & \cite{sinigaglia2020survey} & \makecell[l]{A survey on multi-factor authentication for online\\ banking in the wild}\\
        \midrule
        a18 & \cite{ali2020two} & \makecell[l]{Two-factor authentication scheme for mobile money: A review of\\ threat models and countermeasures}\\
        \midrule
        a22 & \cite{sciarretta2020formal} & \makecell[l]{Formal Analysis of Mobile Multi-Factor Authentication\\ with Single Sign-On Login}\\
        \midrule
        a30 & \cite{jacomme2021extensive} & \makecell[l]{An Extensive Formal Analysis of Multi-factor\\ Authentication Protocols}\\
        \midrule
        a31 & \cite{thomas2021broad} & \makecell[l]{A broad review on non-intrusive active user\\ authentication in biometrics}\\
    \bottomrule
    \end{tabular}
    \end{adjustbox}
    \label{tab:biblio_portfolio}
\end{table}

\section{The Analysis Lenses Application}
\label{S:2.3}

The following lenses were selected in this work: (a) the perspective of the threats analyzed in each work and (b) the characteristics of each model/scheme proposed in the articles listed in the portfolio. The threat analysis (a) aims to explain the main points of vulnerability within the authentication currently used, which deserve greater care during the design and construction of an authentication mechanism. On the other hand, characteristics (b) are associated with additional functionalities or problem-solving results in each work in the portfolio.

We will start with the work of Cheng, Lee, and Hsu \cite{chen2012mobile} who propose a lightweight user authentication scheme, which uses few resources, adopts \textit{hash} functions, and proposes to have an integration between the biometrics of a device mobile for authentication in systems. Its authentication scheme consists of four phases: registration, \textit{login}, authentication, and password change. Its threat model aims to address the following threats: \textit{insider attack}, \textit{stolen-verifier attack}, \textit{impersonation attack}, \textit{replay attack}, \textit{reflection} and \textit{parallel session attack}, \textit{denial-of-service attack}, and \textit{password guessing attack}.

In Li \textit{et al.} \cite{li2014unified} a unified model is proposed that derives its operation from a threat tree. According to the author, its performance is superior to traditional threat trees. Its major gains are mitigating threats in a cheaper way and with mitigations already cataloged in this new proposed model. Classification through STRIDE and threat representation through a DFD is also used as process tools.

As for Dhillon and Kalra \cite{dhillon2017secure}, the solution is a multi-factor and mutual authentication. Proposing to be a robust and lightweight authentication, it uses XOR, and \textit{hash} functions for the authentication protocol even though it uses mutual authentication. It brought as desired security features mutual authentication, confidentiality, user anonymity, availability, \textit{forward secrecy}, scalability, and attack resistance. Also, the same work lists in its attack model the following items: \textit{eavesdropping attack}, \textit{impersonation attack}, \textit{man-in-the-middle attack}, \textit{denial of service attack}, \textit{parallel session attack}, \textit{password change attack}, \textit{gateway node bypassing attack}, and \textit{offline guessing attack}; The proposal takes place in four phases: registration, \textit{login}, authentication, and password change.

Ferrag, Maglaras, and Derhab \cite{ferrag2019authentication} presented an important contribution to a discussion about unusual authentication factors. Some factors can be listed, such as touch dynamics, rhythm, ear shape, and arm gestures. The psychological and behavioral factors are also discussed in the work. His contribution is based on a large amount of work-related to different types of authentication. It brings to light this evolution in the acquisition and processing of human signals, albeit subjective in some cases.

Sinigaglia \textit{et al.} \cite{sinigaglia2020survey} provides a comprehensive survey on technologies and challenges of using multi-factor authentication, specifically for banks (financial system). Although its analysis took place from the perspective of user authentication and not of devices, some evaluations can be ported between the two. As a threat model, a model is presented with the following threats listed: device theft, duplicate authenticator, shoulder surfer, \textit{eavesdropping software}, \textit{social engineer}, \textit{man-in-the- browser}, and \textit{man-in-the-mobile}. In general, their threats focus more on intended objectives and not on the steps needed to consolidate the attack, as in other classic models.

Ali, Dida, and Sam \cite{ali2020two} provide a deeper analysis of authentication schemes for \textit{mobile money}. This work focuses on user authentication for payments through mobile devices using two-factor authentication. The authors divide the threats in their model into five groups: attacks against privacy, attacks against authentication, attacks against confidentiality, attacks against integrity, and attacks against availability. For the present work, only the attack group against authentication was brought, consisting of: \textit{impersonation attacks}, \textit{replay attacks}, \textit{masquerade attack}, \textit{spoofing attack}, \textit{social engineering attack}, \textit{phising attack}, and \textit{trojan horse attack}. This work also focuses on technologies for user authentication.

Sciarretta \textit{et al.} \cite{sciarretta2020formal} provide a formal analysis of an MFA with Single Sign On (SSO) where two \textit{e-health} scenarios are used to support the analysis. As factors, a \textit{token authentication} through One Time Password (OTP) and a Challenge-Response (CR) is used. As threats, the following are brought up: device thief \textit{smartphone}, device thief and IDCard, social engineering, shoulder surfer, App duplicator, leaking \textit{software}, and malicious application. It can be noted that the analyzed proposal aims at user authentication, and due to this, some of the threats are directly addressed to problems with human users.

Thomas and Mathew \cite{thomas2021broad} also showed an approach that considers behavior-based authentication factors. As an example of authentication factors used in this article, we can mention bio-signals, emotion recognition, and typing pattern. Therefore, the authors comprehensively review non-intrusive active methods for user authentication. Still, his work has presented the importance of non-intrusive methods for authentication, which can be the basis of continuous and active authentication in future research on computer systems.

Finally, there is the work of Jacome and Kremer \cite{jacomme2021extensive}, which proposes a formal analysis of multi-factor authentication. In this work, the authors evaluate Google 2-steps and FIDO's U2F through formal methods using \textit{applied pi-calculus} and the Proverif tool. The work also presents a threat model composed of the following threats: compromised passwords, network control, compromised platform, human aspects, and "trust this computer mechanism" - threats very specific to the analyzed MFA models.

\begin{table}[tp]
    \centering
    \caption{Correlation between the set of attacks cited and the articles in the bibliographic portfolio. In the lines is listed the set of attacks, the third column classifies the attack ((C)onfidentiality, (I)ntegrity, and (A)vailability), and the other columns show the articles where such attacks are mentioned. Also, $\bullet$ means present in the article, and $\circ$ means not present. Further, not all articles appeared in the columns since some did not present a threat model.}
     \begin{adjustbox}{max width=.85\textwidth}
    \begin{tabular}{p{0.05\textwidth}p{0.25\textwidth}p{0.05\textwidth}p{0.1\textwidth}p{0.1\textwidth}p{0.1\textwidth}p{0.1\textwidth}p{0.1\textwidth}}
        \toprule
        \#& Attacks & CIA & \cite{chen2012mobile} & \cite{dhillon2017secure} & \cite{sinigaglia2020survey} & \cite{sciarretta2020formal}
        &\cite{ali2020two}\\
        \toprule
        01& Insider &C               &$\bullet$&$\circ$&$\circ$&$\circ$&$\circ$\\
        02& Stolen-Verifier &C        &$\bullet$&$\bullet$&$\bullet$&$\circ$&$\circ$\\
        03& Impersonation &C         &$\bullet$&$\bullet$&$\circ$&$\circ$&$\bullet$\\
        04& Replay &I
        &$\bullet$&$\circ$&$\circ$&$\circ$&$\bullet$\\
        05& Reflection &C            &$\bullet$&$\circ$&$\circ$&$\circ$&$\circ$\\
        06& Parallel Session &C       &$\bullet$&$\bullet$&$\circ$&$\circ$&$\circ$\\
        07& Denial-of-Service &A &$\bullet$&$\bullet$&$\circ$&$\circ$&$\circ$\\
        08& Password Guessing &C     &$\bullet$&$\bullet$&$\circ$&$\circ$&$\circ$\\
        
        09& Eavesdropping &C         &$\circ$&$\bullet$&$\bullet$&$\bullet$&$\circ$\\
        10& Man-In-The-Middle &C     &$\circ$&$\bullet$& $\bullet$ &$\circ$&$\circ$\\ 
        11& Password Change &C       &$\circ$&$\bullet$&$\circ$&$\circ$&$\circ$\\ 
        12& Gateway Node Bypassing &C &$\circ$&$\bullet$&$\circ$&$\circ$&$\circ$\\
        
        13& Duplicate Authenticator &C &$\circ$&$\circ$&$\bullet$&$\circ$&$\circ$\\ 
        14& Shoulder Surfer &C       &$\circ$&$\circ$&$\bullet$&$\bullet$&$\circ$\\
        15& Social Engineer &C       &$\circ$&$\circ$&$\bullet$&$\bullet$&$\bullet$\\ 
        16& Man-In-The-Browser &C    &$\circ$&$\circ$&$\bullet$&$\circ$&$\circ$\\ 
        17& Man-In-The-Mobile &C      &$\circ$&$\circ$&$\bullet$&$\circ$&$\circ$\\
        
        18& App Duplicator &C        &$\circ$ &$\circ$&$\circ$&$\bullet$&$\circ$\\
        19& Malicious Application &C  &$\circ$ &$\circ$&$\circ$&$\bullet$&$\circ$\\
        
        20& Masquerade &C            &$\circ$&$\circ$&$\circ$&$\circ$&$\bullet$\\
        21& Spoofing &C              &$\circ$&$\circ$&$\circ$&$\circ$&$\bullet$\\    
        22& Phishing &C                &$\circ$&$\circ$&$\circ$&$\circ$&$\bullet$\\
        23& Trojan Horse &I           &$\circ$&$\circ$&$\circ$&$\circ$&$\bullet$\\
        \midrule
    \end{tabular}
    \end{adjustbox}
    \label{tab:slr_threat_model_relateds}
\end{table}

\begin{comment}
%analise sobre uso de autenticação mutua
Although some so-called lightweight protocols that use mutual authentication for user identification add confidentiality to the process, adding functionality overloads the authentication process and makes it more expensive than necessary. Attach features and requirements that directly address data transmission.
\end{comment}

The attacks listed in the Table (\ref{tab:slr_threat_model_relateds}) have a wide range of impact, target audience, and security domain. As with DDoS, some can have major financial impacts on institutions and systems. Others have the target audience focused on specific users or devices, and the impact is associated with the level of clearence of that person in the system. As for the security domain, some attacks use technology to attack the network, the fragility of a chosen password; however, others focus on less technological issues, as is the case with shoulder surfer and \textit{social engineering}. In this way, it is important to know the possible threats to the system, as explained in the portfolio.

\section{Results and Discussion}
\label{S:3}

\begin{figure}[b]
    \centering
    \includegraphics[width=.65\textwidth]{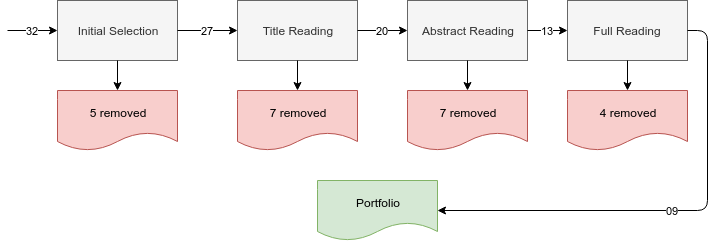}
    \caption{Systematic Literature Review Workflow - the diagram shows in gray the processes, in red the removed documents, and in green the resulting portfolio.}
    \label{fig:slr_workflow}
\end{figure}

Through this work, it was possible to answer the three initial questions (Q1, Q2, and Q3) that motivated him. During its development, an SLR was created, Figure \ref{fig:slr_workflow}, nine articles were selected as the final portfolio from an initial selection of 32 articles using the Table \ref{tab:criteria_inc_exc} as criteria for initial document selection and the query-string (\ref{eq:query_slr}) during the search.

As for Q1, a portfolio of nine articles was evaluated in detail and represents the state-of-the-art at the intersection of threat model and multi-factor authentication areas. During the process, the selected criteria helped to avoid bias and improve assertiveness in the \textemdash article selection, also helping in the reproducibility of this work. Also, the SCOPUS database proved sufficient as it incorporated several computer science sources and publishers. Additionally, selecting tools, variables, and spreadsheet templates made the work less difficult.

\begin{figure}
     \centering
     \subfigure[Articles Categorization]{
         \includegraphics[width=0.65\linewidth]{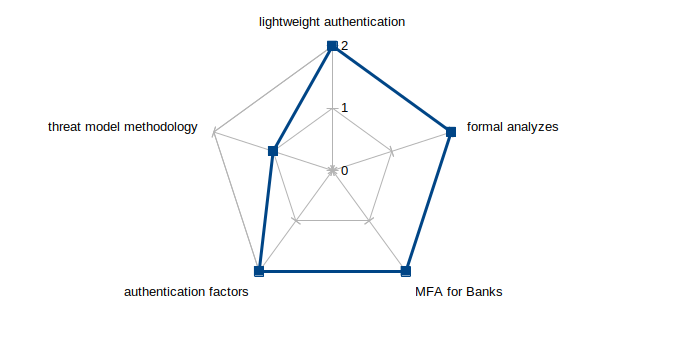}
     }
     %\hfill
    % \subfigure[Occurrences of attacks]{
    %    \includegraphics[width=0.4\linewidth]{figs/pizza_ocorrencias.png}
    %}
    %\hfill
    % \subfigure[Attack Classification]{
    %    \includegraphics[width=0.3\linewidth]{figs/pizza_classificacao.png}
    %}
    \caption{Quantitative data obtained from the analysis of portfolio articles}
    \label{fig:portfolio_data}
\end{figure}

As for the characteristics (Q2) of FAC works, we can categorize them (Figure \ref{fig:portfolio_data}-a) into five different ones: lightweight authentication (2 articles), threat model methodology (1 article), authentication factors (2 articles), MFA for Banks (2 articles), and formal analyzes (2 articles). From these categories, it is possible to see that there is an effort for authentication to evolve to a lighter, more diversified form and that its processes are unquestionably validated through formal methods. Therefore, there is also a concern about its use in financial institutions and how to assess its threats at a lower cost. In general, low cost and resource restrictions are key factors in multi-factor authentication and threat model research.

Finally, regarding threats (Q3), four documents were used for their listing (see Table \ref{tab:slr_threat_model_relateds}). Among the threats listed, none appeared in all the works, and they are distributed (Figure \ref{fig:portfolio_data}-b) in three occurrences (17.3\%), two occurrences (21.7\%), and one occurrence (60.8\%). A total of 23 distinct threats were listed in this survey, and they range (Figure \ref{fig:portfolio_data}-c) from confidentiality threats, which have the most items (20 threats), integrity (2 threats), and availability (1 threat) \textemdash with the fewest. Consequently, we notice that some types of threats appear more frequently, such as confidentiality, which is the category of threats with the highest occurrence, adding up to 86.95\% of the works listed in Table \ref{tab:slr_threat_model_relateds}. This result is because threats linked to authentication are directly associated with reliability, which involves the privacy, information disclosure, and secrecy - not detailed in the CIA-based classification of Stallings\cite{stallings2012computer}.

In brief, we can say that by answering these three questions, it was possible to have a good representation of the state-of-the-art and the paths that research in multi-factor authentication and threat modeling has taken. Furthermore, textual items (tables and charts) and graphical items (schemes, graphs, and images) provide a set of artifacts that support future research or updates of this work.

\section{Conclusion and Future Works}
\label{S:4}

This work fully achieved its objectives by listing through the portfolio the main articles within the research area, bringing the main characteristics of solutions through the portfolio analysis, and presenting the main threats found in the related literature. Also, it provides a quantitative analysis of the portfolio article content, discuss and comment this, and report it in Table \ref{tab:slr_threat_model_relateds} and Figure \ref{fig:portfolio_data}.

It is important to monitor the area and the evolution of research in threat models and multi-factor authentication in future work. Also, backward and forward snowballing processes must improve the specific knowledge acquired in each category. Therefore, an important evolution of this work is the refinement of the search through terms that focus on specific technologies such as 6G, Fog Computing, or continuous authentication, and there are growing research trends.

%%%%%%%%%%%%%%%%%%%%%%%%%%%%%%%%%%%%%%%%%%%%%%%%%%%%%%%%%%
%%%%%%%%%%%%%%%%%% Referencias
%%%%%%%%%%%%%%%%%%%%%%%%%%%%%%%%%%%%%%%%%%%%%%%%%%%%%%%%%%

\bibliographystyle{cas-model2-names}

\bibliography{cas-refs}

\end{document}